\documentclass[11pt,twoside]{article}
\usepackage{asp2004}
\usepackage{psfig}
\usepackage{epsf}
\usepackage{graphicx}
\usepackage{lscape}
\begin{document}
\title{
The Secular Decrease of the Crab Nebula at 927 and 151.5 MHz
}
\author{
E.\,N.\,Vinyajkin
}
\affil{
Radiophysical Research Institute (NIRFI),\\
25 B.\,Pecherskaya st.,
Nizhny Novgorod,  603950, Russia
}

\begin{abstract}
Long-term measurements have been carried out of the Crab Nebula
radio emission flux  density relative to Orion A at 927~MHz
and relative to Cygnus~A and Virgo~A at 151.5~MHz.
As a result the mean rates have been found of the
secular change of the Crab Nebula radio  emission flux density:
$d_m\,(927\hbox{ MHz})=-(0.18\pm0.10)$\%\,year$^{-1}$
(over the period 1977--2000), $d_m\,(151.5\hbox{ MHz})=-
(0.32\pm0.08)$\%\,year$^{-1}$ (over the period 1980--2003).
\end{abstract}

\thispagestyle{plain}

\section{Introduction}

The Crab Nebula (the supernova 1054 remnant, radio source Taurus~A)
shows relatively slow evolution of radio emission which was observed
at decimeter and centimeter wavelengths (see the Table).
Since 1977 measurements of the Crab Nebula flux density relative to
Orion~A have been carried out at
the NIRFI Radioastronomical Observatory ``Staraya Pustyn$'$" at 927~MHz
(Vinyaikin (1993)).
To find a possible dependence of the Crab Nebula radio emission
secular decrease rate on frequency measurements were also
carried out relative to Cygnus~A and Virgo~A at 151.5~MHz.

\section{Observations}

{\bf 927 MHz.} The measurements of the Crab Nebula flux density
$S^{\rm Crab}$ relatively to Orion~A $S^{\rm Orion~A}$ have been
carried out at 927~MHz with the 10-m radio telescope. Figure~1
and the Table show the measurement results of the ratio
${(S^{\rm Crab}/S^{\rm Orion~A})}_{927\hbox{ MHz}}$  in the interval
1977--2000. The solid straight line is a weighted
least-squares fit and a horizontal dashed line shows a weighted
average value over all measurements. It is possible that the
Taurus~A flux density decrease is not uniform. There was a
relatively quicker decrease between 1983 and 1989. The mean value of
the Crab Nebula radio emission secular decrease rate
$d(\nu,t)={(S^{\rm Crab})}^{-1}\,dS^{\rm Crab}/dt$  equals to
$d_m\,(927\hbox{ MHz})=-(0.18\pm0.10)$\%\,year$^{-1}$ over the period
1977--2000. Figure~2 shows the results of the parallel measurements
of the Cygnus~A and the Orion A flux densities ratio
at 927~MHz. The solid straight line is again a weighted 
least-squares fit and a horizontal dashed line shows a weighted average
value over all measurements. Practically, these lines coincide.
The absolute value of ``the secular change" mean rate is
$|d_m|<0.04$\%\,year$^{-1}$
over the same period 1977--2000. Thus, these measurements
confirm the stability of the Cygnus~A and the Orion A radio
emission.

\vspace{-8mm}

\begin{figure}[ht]

\begin{minipage}[t]{65mm}
\center
\includegraphics[width=65mm]{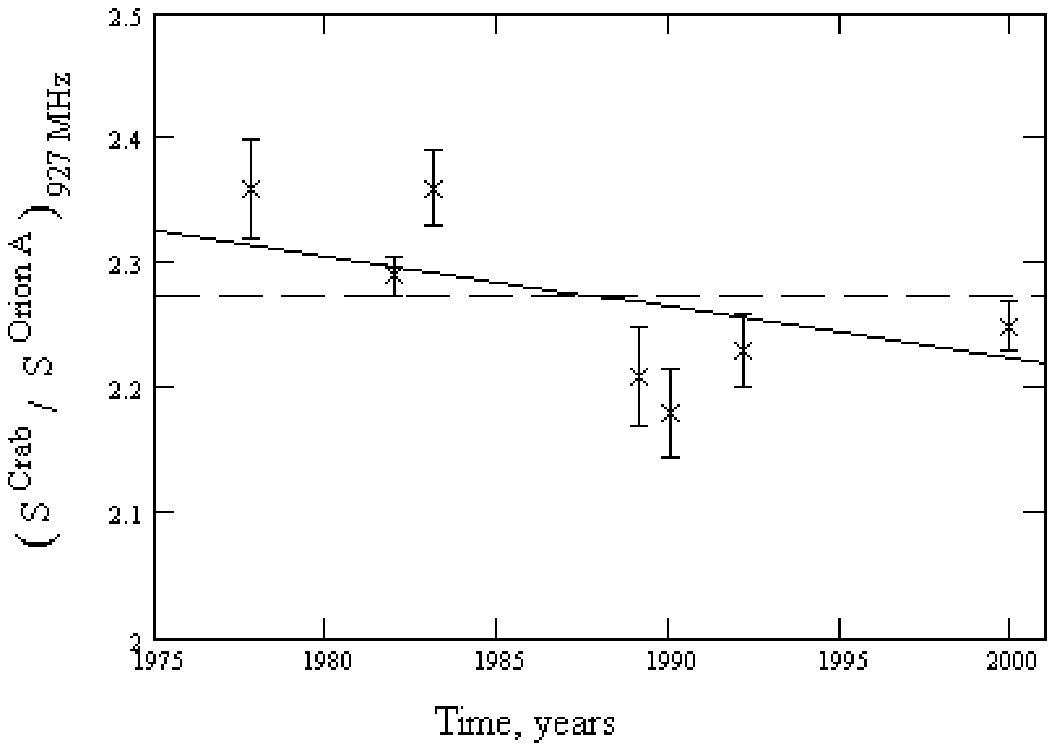}
\vspace{-15mm} \caption[]{The Crab Nebula and Orion~A  flux
densities ratio at 927~MHz versus time } \label{Fig1}
\end{minipage}
\begin{minipage}[t]{65mm}
\center
\includegraphics[width=64mm]{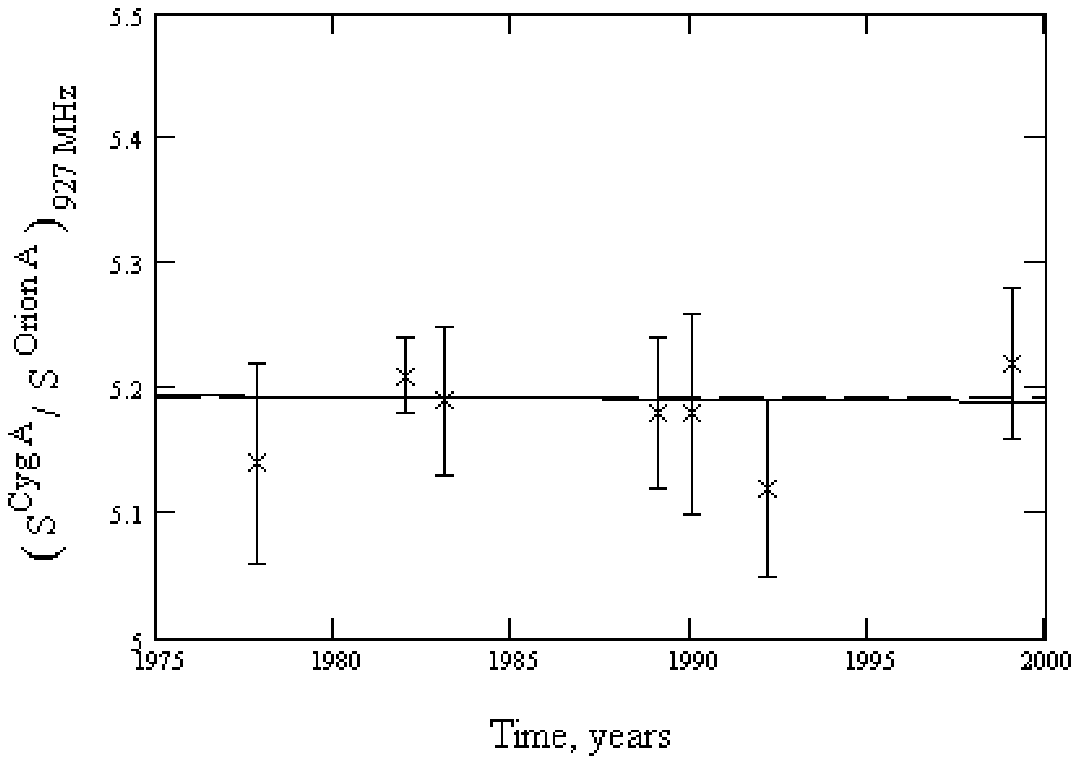}
\vspace{-15mm} \caption[]{Cygnus~A and Orion~A flux densities ratio
at 927~MHz versus time } \label{Fig2}
\end{minipage}
\end{figure}

\noindent
\begin{tabular}{|c|c|c|c|}
\multicolumn{4}{l}{{\bf Table.} \ The Taurus A secular decrease rates}\\
\hline
Frequency,  & Period,  & The mean secular decrease &
Reference\\
MHz &  years &  rate $d_m$, \%\,year$^{-1}$  & \\
\hline
927 & 1962--1977 & $-0.18\pm 0.01$ & Vinyajkin \& Razin \\
 &  & &  (1979)\\
\hline
927 & 1977--2000 & $-0.18\pm0.10$ & This report\\
\hline
8000& 1968--1984 & $-0.167\pm 0.015 $& Aller \& Reynolds \\
&  & &  (1985)\\
\hline
151.5 & 1980--2003 & $-0.32\pm0.08$& This report\\
\hline
\end{tabular}

\vspace{6mm}

{\bf 151.5 MHz.} The measurements relative to Cygnus~A and Virgo~A
at 151.5~MHz were made at the
``Staraya Pustyn$'$" with a radio interferometer consisting of two
identical 14-m radio telescopes with a base line of about 60~m.
As a result
of the measurements it has been found that
$d_m\,(151.5\hbox{ MHz})=-(0.32\pm0.08)$\%\,year$^{-1}$
over the period 1980--2003.
Measurements of
the ratio of Cygnus~A and Virgo~A radio emission flux densities
at 151.5~MHz were also carried out for the same epochs.
These measurements showed the stability of the ratio at the level
less than $0.1$\%\,year$^{-1}$ by the absolute value.





\section{Discussion}

It is seen from the Table  that the value of $d$\,(927~MHz)
averaged over a large periods of time was constant in the
limits of errors during the last 40 years. In the Table there is a
hint on a more quicker decline of the Crab Nebula radio emission at
2~m wavelength as compared with decimeter and centimeter wavebands.
To make such a conclusion, however, we need further observations.

{\bf Acknowledgment.} This work has been supported by the
International Science and Technology Center under the ISTC project
No.\,729.

\end{document}